\documentstyle[12pt]{article}
\begin{document}
\pagestyle{empty}
\rightline{UG-4/93}
\rightline{July 1993}
\vspace{2.0truecm}
\centerline{\bf  A BRST Analysis of $W$-symmetries}
\vspace{1.5truecm}
\centerline{E.~Bergshoeff, H.J.~Boonstra, S.~Panda\footnote{address after
 August 1 1993: The Mehta Research Institute of Mathematics and
 Mathematical Physics, 10 Kasturba Gandhi Marg, Allahabad, 211002 India}
  and
M.~de Roo}
\vspace{.5truecm}
\centerline{Institute for Theoretical Physics}
\centerline{Nijenborgh 4, 9747 AG Groningen}
\centerline{The Netherlands}
\vspace{1.5truecm}
\centerline{ABSTRACT}
\vspace{.5truecm}
We perform a classical BRST analysis of the symmetries corresponding to a
generic $w_N$-algebra. An essential feature of our method is
that we write the $w_N$-algebra in a special
basis such that the algebra manifestly has a ``nested'' set of
subalgebras
$v_N^N \subset v_N^{N-1} \subset \dots \subset v_N^2 \equiv w_N$
where the subalgebra
$v_N^i\ (i=2, \dots ,N)$ consists of generators
of spin $s=\{i,i+1,\dots
,N\}$, respectively. In the new basis the
BRST charge can be written as a ``nested'' sum of $N-1$ nilpotent
BRST charges.

In view of potential applications to (critical
and/or non-critical) $W$-string theories we discuss the quantum extension
of our results. In particular, we present the
quantum BRST-operator for the $W_4$-algebra in the new basis.
For both critical and
non-critical $W$-strings we apply our results to
discuss the relation with minimal models.

\vfill\eject
\pagestyle{plain}

\section{Introduction}

In recent years it has turned out that in order to describe string
theory it is convenient to use the BRST formalism \cite{bec1}.
For instance, via a BRST analysis one can derive the critical dimension
and calculate the spectrum of the theory.
For critical strings this was first done in \cite{ka1}.
More recently, the spectrum of non-critical strings has been
calculated using this formalism \cite{li1}. The BRST approach also
plays a crucial role in the construction of a string field theory
\cite{si1}.

The starting point in the BRST approach is the introduction of a set
of canonical variables (the ``string coordinates'') which satisfy a
standard Poisson bracket. In string theory the relevant variables are
given by a set of holomorphic variables and a set of anti-holomorphic
variables. We restrict the BRST analysis to
the holomorphic sector since the two sectors require a similar treatment.
The two-dimensional conformal symmetries of string theory
are encoded in a set of first-class constraints on the string coordinates
whose
Poisson brackets are given by the Virasoro algebra. Given this
Virasoro algebra one can construct a nilpotent BRST charge by
extending the phase space with a set of anticommuting ghost variables.
At the classical level, this BRST charge can be used to define the
physical variables of the theory. In a canonical quantization the
Poisson brackets get replaced by
so-called Operator Product Expansions (OPE's) where the operators act
in a Hilbert space.
At the same time the BRST charge gets replaced by a nilpotent
BRST operator.
The physical states in the Hilbert space are defined as the cohomology
classes of this BRST operator. The BRST operator thus provides a
convenient way to calculate the spectrum of the theory.

Due to normal ordering problems it is not guaranteed {\sl a priori}
that a nilpotent BRST operator can be constructed. If this is not the
case one cannot define the physical states and the theory is said to be
anomalous. In most cases the BRST operator can be made nilpotent provided
that certain conditions hold. For instance, in the case of the bosonic
critical string requiring nilpotency of the BRST operator leads to
the condition that the number of string coordinates is 26, i.e.
the bosonic string moves in a 26-dimensional spacetime \cite{ka1}.

Within the BRST formalism it is rather natural to extend the Virasoro
constraints with a set of additional
first class constraints and investigate whether this extended set
still leads to a sensible spectrum thus providing the basis for
the construction of new string theories \cite{bi1}.
The complete set of
first-class constraints must form a closed Poisson-bracket algebra
which is an extension
of the classical Virasoro algebra\footnote{A few clarifying remarks concerning
the terminology ``classical'' algebras are
in order here. In general, by a classical algebra is meant a Poisson-bracket
algebra.
In this sense there exist a classical Virasoro algebra with a so-called
central extension. However, in this paper we will always reserve the
term ``classical'' algebra for the special case where this central
extension is zero. For the realization of the Virasoro algebra in terms of free
fields this means that we do not consider background charges at the
classical level. Similarly, by a classical $w_N$-algebra (see below)
we mean a Poisson
bracket algebra whose free field realization contains only {\sl single}
derivatives of the fields.}. Most of the recent research has focussed
on algebras where the new generators carry a spin which is higher than
the spin of the Virasoro generators. Such algebras are
denoted as extended conformal algebras or, briefly,
``$w_N$-algebras'' where $N$ indicates the highest spin
of the generators involved (usually one uses a convention in which the
Virasoro generators carry a spin equal to two).
The simplest example, which has been
mostly studied, is the $w_3$-algebra which involves the Virasoro
generators and a generator of spin three \cite{za1}. The
$w_3$-algebra is quadratically nonlinear, i.e., Poisson brackets
of the constraints lead to polynomials of the constraints which are
at most quadratic. The BRST charge of the $w_3$-algebra
was first constructed in \cite{thm1} while the  BRST charge for
general quadratically nonlinear algebras was obtained in \cite{sc1}.

In view of potential applications to $W$-string theories it is necessary to
quantize the $w_N$-symmetries via the BRST formalism and to perform
a spectrum analysis.
One noteworthy feature that has emerged from this quantisation
is that although classically the first-class
constraints always form a closed Poisson bracket algebra, the corresponding
quantum operators do not necessarily form a closed quantum algebra in the
full Hilbert space,
even after including possible renormalizations of the generators
and allowing for quantum deformations of the classical algebra\footnote{
To be more precise, the existence of a quantum algebra depends on the
basis one is using for the classical algebra. Using the standard,
so-called Miura (see below), basis of the $w_N$ algebra, there exists
a corresponding quantum algebra which we denote by $W_N$. This is however
not the case if we use our new, realization-dependent, basis
of the $w_N$ algebra (see below).}.
Indeed, they do not have to form a closed quantum algebra. All one
needs in the BRST approach at the quantum level is the existence of
a nilpotent BRST operator. So we have the following picture:

\begin{eqnarray}
{\rm classical} &\rightarrow& {\rm closed\ Poisson\ bracket\ algebra}
\nonumber\\
{\rm quantum} &\rightarrow& {\rm nilpotent\ BRST\ operator}
\end{eqnarray}
A recent example of a nilpotent BRST operator without a corresponding quantum
algebra was given in \cite{ber1}. In the present work we will encounter
more examples. Once a nilpotent BRST operator has been constructed,
its cohomology, and hence the spectrum of the theory, can be computed.
The quantum constraints, which by construction are BRST-trivial,
then close within the space of cohomology classes of the BRST operator.

It is the purpose of this paper to give a systematic BRST analysis
of general $w_N$-symmetries both at the classical as well as at the quantum
level. Sofar, explicit results are known and well
understood only in the case of the $w_3$-algebra.
In \cite{horn,zhu2},
an expression has been presented for the BRST operator of the $w_4$-algebra.
However, the complexity of this expression makes it rather hard to
deal with in practice.
Recently it has been pointed out that in
case of the $w_3$ algebra the BRST analysis can be simplified by making
an appropriate redefinition of the canonical variables \cite{po1}.
After the redefinition the BRST charge can be written as the
sum of two charges that are separately nilpotent. It is expected that
this will lead to simplifications in the analysis of the spectrum in
the quantum case.
In \cite{be1} the redefinition of the canonical variables was
translated into a corresponding redefinition of the
generators and it was indicated how a similar
simplification could be made for the generic $w_N$-algebra. The
additional structure which arises after the redefinitions makes
it possible to obtain a relatively simple structured expression
for the BRST-operator for $W_4$ (see Section 5), and
in principle also for $W_N$.

The general picture that arises and which is confirmed by the present
work is as follows. Usually the $w_N$-algebra is realized in terms of $N-1$
free scalar fields and given in a special basis
which is related to making a so-called Miura transformation. We will
call this special basis the ``Miura basis''. In this Miura basis the
BRST charge of the $w_N$-algebra is a rather complicated expression
which for growing $N$ contains terms of increasingly high order
in the ghost fields. For instance, the BRST charge of the $w_3$-algebra
is at most trilinear in the ghosts but the BRST charge of the $w_4$-algebra
(see Section 4) contains already terms of seventh order in the ghosts.
In the next section we will show how the generators of the
$w_N$-algebra can be redefined such that the $w_N$-algebra contains a
``nested'' set of subalgebras
\begin{equation}
\label{eq:nesal}
v_N^N \subset v_N^{N-1} \subset \dots \subset v_N^2 \equiv w_N
\end{equation}
where the subalgebra
$v_N^i\ (i=2, \dots ,N)$ consists of $(N-i+1)$ generators
$\{w_N^{i}, w_N^{i+1},\dots ,w_N^{N}\}$ of spin $s=\{i,i+1,\dots
,N\}$, respectively. The generators are
realized by $N-1$ free (holomorphic) scalar fields
$\phi_n, n=1,\dots,N-1,$
such that the generator $w_N^N$, of highest spin, only
depends on the single scalar $\phi_{N-1}$, the generator $w_N^{N-1}$, of next
to highest spin, only depends on the two scalars $\phi_{N-1},\phi_{N-2}$,
etc. Finally,
the Virasoro generator $w_N^2$ is the only generator that depends on
{\sl all} scalars $\phi_1, \dots ,\phi_{N-1}$.
This particular dependence of the
generators on the scalars automatically leads to the nested
subalgebra structure indicated in (\ref{eq:nesal}). For instance, since
the highest spin generator $w_N^N$ only depends on $\phi_{N-1}$, and all
other generators contain other scalars as well, the Poisson bracket
algebra of $w_N^N$ must close on itself thus leading to the subalgebra
$v_N^N$ etc.

An immediate consequence of the new basis is that the scalar $\phi_1$
which only occurs in the Virasoro generator can be replaced there by a
term $\partial X^\mu\partial X_\mu$ containing an
arbitrary number of scalars $X^\mu$ without upsetting the closure of the
algebra since this term commutes with all the other generators. This leads to a
multi-scalar realisation of the $w_N$-algebra. Such multi-scalar realisations
were first considered in \cite{ro1}.
The above structure is summarized schematically in Table 1.

\begin{center}
\begin{tabular}{|l||l|l|l|c|l|}
\hline
\hfil  &\hfil&\hfil&\hfil&\hfil  &\hfil\\
$w_N^2$&$X^\mu$&$\phi_2$&$\phi_3$&$\dots$&$\phi_{N-1}$\\
$\vdots$&$\vdots$&$\vdots$&$\vdots$&$\dots$&\\
$\vdots$&$\vdots$&$\vdots$&$\vdots$&$\dots$&\\
$w_N^{N-2}$&$\phi_{N-3}$&$\phi_{N-2}$&$\phi_{N-1}$&&\\
$w_N^{N-1}$&$\phi_{N-2}$&$\phi_{N-1}$&&&\\
$w_N^N$&$\phi_{N-1}$&&&&\\
\hfil  &\hfil&\hfil&\hfil&\hfil  &\hfil\\
\hline
\end{tabular}
\end{center}

\vspace{.5truecm}

\noindent {\bf Table\ 1.} \ \ \ \ \ This table shows the generic structure
of the $w_N$-algebra in the new basis discussed in Section 2.
The left column indicates the
generators $\{w_N^2,\dots ,w_N^N\}$ of the algebra. The other colums indicate
the dependence of the generators on the scalars $\{X^\mu,\phi_2,
\dots ,\phi_{N-1}\}$.

\vspace{.5truecm}

In order to construct the BRST charge of the complete $w_N$-algebra one can
now first consider the smallest subalgebra $v_N^N$ generated by $w_N^N$.
Its corresponding BRST charge we denote by $Q_N^N$. One then considers the
next subalgebra $v_N^{N-1}$ generated by $\{w_N^N,w_N^{N-1}\}$ which
has its own BRST charge $Q_N^{N-1}$. Since $v_N^N \subset v_N^{N-1}$
we have that $Q_N^N \subset Q_N^{N-1}$. By this we mean that if one sets
the ghost variables corresponding to the spin $N-1$ symmetries equal
to zero the expression for $Q_N^{N-1}$ equals that of $Q_N^N$. In general,
this does not imply that the BRST charge  $Q_N^{N-1}$ can be
written as $Q_N^{N-1} = Q_N^N$ + ``rest'' such that the ``rest'' terms
are separately nilpotent. The fact that this does happen for the
$w_3$-algebra is an exception (see below). We thus arrive at the
following ``nested'' structure of the BRST charge $Q_N$ of the
$w_N$-algebra:
\begin{equation}
\label{eq:Qnest}
Q_N^N \subset Q_N^{N-1} \subset Q_N^{N-2} \subset \dots \subset
Q_N^3 \subset Q_N^2 \equiv Q_N
\end{equation}
Here the inclusion symbols indicate how the different (nilpotent) BRST charges
can be obtained from each other by setting certain ghost variables
equal to zero. A nice feature of this structure is that one can investigate
systematically the BRST charges of the different nested subalgebras
and thus iteratively construct the BRST charge of the complete $w_N$-algebra.
In this paper we will present results for the subalgebras $v_N^N$ and
$v_N^{N-1}$ for {\sl any} $N$.

Note that  the generator
 $w^2_{N}$ always satisfies by itself the Virasoro algebra. Therefore
 there is, besides the nested structure (\ref{eq:Qnest}), also
 a nilpotent BRST charge corresponding to this Virasoro subalgebra.
 This BRST charge is in fact given by $Q^2_N-Q^3_N$. This is
 the reason that for $N=3$ the  nested structure (\ref{eq:Qnest})
 is given by
\begin{equation}
Q_3^2 = Q_0 + Q_1
\end{equation}
where $Q_0=Q^2_3-Q^3_3$ and
 $Q_1=Q^3_3$ are two anti-commuting nilpotent BRST-charges
\cite{po1,be1}.

It is to be expected that the nested structure (\ref{eq:Qnest}) of
 the BRST-charges
 survives quantisation\footnote{To distinguish between classical and
quantum expressions, we will write the quantum expressions with boldface.}.
The examples given in this paper provide
 arguments in favour of this conjecture. In that case the nested structure
discussed in this paper
 should be useful in the construction of the spectrum of the
 $W_N$-string.

In \cite{ddr,rama}, a relationship was suggested between
the spectra of $W_N$-strings and Virasoro minimal models.
In the case of the
$W_3$-string this relation has been made more explicit in
\cite{po1,po2,po3,we1,hu1}.
In particular, it was
shown that the $W_3$-string can be viewed as an
ordinary $c=26$ string, where the matter sector includes a
$c={\textstyle{1\over 2}}$ Ising model.
{}From the point of view of the nested structure (\ref{eq:Qnest}), it is easy
to see how the
$c={\textstyle{1\over 2}}$ Ising model enters into the game by
observing the following numerology. Since
the $v_3^3$ subalgebra has its own nilpotent BRST operator ${\bf Q}_3^3$,
one can separately construct its cohomology. The BRST operator ${\bf Q}_3^3$
is realized
by a single free scalar $\phi_2$ and the ghosts of the spin-three symmetries.
It turns out that the total central charge $c_3^3$ of these fields equals
${\textstyle{1\over 2}}$ which is precisely that of the Ising model.
In this paper we will apply a similar numerology to the nested structure
of a generic $w_N$ algebra. Our results suggest a very general relationship
between the spectra of $W_N$ strings and $W$ minimal models.
A similar relationship is
suggested between the so-called non-critical $W_N$-strings and
$W$ minimal models, thereby extending a conjecture made in \cite{be1}.
It will be interesting to see whether the conjectures will be
confirmed by explicit calculations of the spectra of (critical and/or
non-critical) $W_N$-strings. We hope that the nested structure discussed in
paper will considerably facilitate this task.

The organisation of this paper is as follows.
In Section 2 we show how the redefinition of the $w_N$-algebra discussed
above can be carried out
for arbitrary $N$. In Section 3 we present general results for any $N$
for the first two subalgebras $v_N^N$ and $v_N^{N-1}$.  In Section 4 we
discuss the special cases $N=3,4,5$.
The discussion of
Sections 2, 3 and 4 is always at the classical level.
In Section 5  we
extend some of our results to the quantum case. For instance, for $N=4$
we give the quantum BRST operator corresponding to the $w_4$-algebra.
Finally, in section 6 we
discuss the relations with (Virasoro) minimal models for both
critical $W_N$-strings and the non-critical $W_N$-strings of \cite{ber1}.

\vspace{.5truecm}

\section{A new basis for the $w_N$-algebra}

In this section we will introduce the new basis for
 the $w_N$-algebra, starting from realisations of the $w_N$-algebra
 obtained from the Miura transformation \cite{fat1}.
The basic result of this section is given by formulae (\ref{eq:highest}),
(\ref{eq:formula})
where we give a closed expression for {\sl all} generators of the
classical $w_N$ algebra in the new basis described in the
introduction.

The Miura
 transformation generates realisations of $w_N$ in terms of $N-1$
 scalar fields $\phi_n,\ n=1,\ldots, N-1$.
 This construction is iterative in the
 sense that the generators of the $w_{N+1}$-algebra can be
 expressed in terms of those of the $w_N$-algebra and one aditional
 scalar field $\phi_N$
 \cite{ddr,po4}. We will denote the generators of $w_N$
 in the Miura basis by $M^l_N$, where $l$ is the spin,
  $2\le l\le N$. The
 iterative structure induced by the Miura transformation reads
\begin{eqnarray}
\label{Miura1}
  M^{l}_{N+1} &=& \sum_{k=0}^{l} a_{l,k}{}^{N+1} \,(B_N)^{l-k}\,M^{k}_N \,,
\end{eqnarray}
where it is assumed that $M^k_N=0$ for $k>N$. $B_n$ represents the
 scalar field $\phi_n$:
\begin{eqnarray}
\label{Bdef}
   B_n &=& {i\over {\sqrt{2n(n+1)}}} \partial\phi_n\,,
\end{eqnarray}
and the coefficients $a$ in (\ref{Miura1}) are given by
\begin{eqnarray}
\label{adef}
 k\le l\ &&
 a_{l,k}{}^{N+1} = (-1)^{l-k} {(N-l+1-N(l-k))(N-k)!\over (N-l+1)!\,(l-k)!}\,,
     \\
 l<k\le N\ && a_{l,k}{}^{N+1} = 0 \,.
\end{eqnarray}
Eq.\ (\ref{Miura1}) generates realisations of the
 classical $w_N$-algebra starting
 from $M^0_0=1,\ M^1_1=0$. Note that in particular (\ref{Miura1})
 then implies that
\begin{eqnarray}
\label{M0}
   M^0_N &=& 1\,,\\
\label{M1}
   M^1_N &=& 0\,,\\
\label{M2}
   M^2_N &=& -\sum_{n=1}^{N-1} {\textstyle{1\over 2}}n(n+1) \,(B_n)^2 \,.
\end{eqnarray}
The standard form of the energy-momentum tensor is then obtained
 as $T=-2M^2_N$.
 To illustrate the Miura basis we give explicit results for the
 generators of $w_N,\ N=2,3,4,5$ in Table 2.

The generators $M^l_N$ at fixed $N$ form a closed Poisson bracket algebra.
 Clearly, this is then also the case for any linear combination of
 the $M^l_N$. The redefinition we will now discuss uses the
 iterative structure (\ref{Miura1}) to simplify the generators by
 making appropriate linear combinations. The aim is to construct a set
 of generators such that the highest spin depends on only one scalar,
 $B_{N-1}$, the next highest spin on two scalars, etc.

\begin{center}
\begin{tabular}{|l||l|}
\hline
     &\\
$N=2$&$M^2_2=-\,(B_1)^2$\\
     &\\
$N=3$&$M^2_3=M^2_2-3\,(B_2)^2$\\
     &$M^3_3=2\,(B_2M^2_2+\,(B_2)^3)$\\
     &\\
$N=4$&$M^2_4=M^2_3-6\,(B_3)^2$\\
     &$M^3_4=M^3_3+2B_3M^2_3+8\,(B_3)^3$\\
     &$M^4_4=3\,(B_3M^3_3-\,(B_3)^2M^2_3-\,(B_3)^4)$\\
     &\\
$N=5$&$M^2_5=M^2_4-10\,(B_4)^2$\\
     &$M^3_5=M^3_4+2B_4M^2_4+20\,(B_4)^3$\\
     &$M^4_5=M^4_4+3B_4M^3_4-7\,(B_4)^2M^2_4-15\,(B_4)^4$\\
     &$M^5_5=4\,(B_4M^4_4-\,(B_4)^2M^3_4+\,(B_4)^3M^2_4+\,(B_4)^5)$\\
     &\\
\hline
\end{tabular}
\end{center}

\vspace{.5truecm}

\noindent {\bf Table\ 2.} \ \ \ \ \ Generators of $w_N$ in the
 Miura basis for some low values of N.

\vspace{.5truecm}

As an example, let us perform this redefinition explicitly for
 the first nontrivial case, $N=4$. We start with the highest
 spin generator, $M^4_4$. As we see in
 Table 2, it depends on $M^3_3$ and $M^2_3$. However, these
 can be expressed in terms of $M^3_4$ and
 $M^2_4$ by inverting the relations given in Table 2:
\begin{eqnarray}
\label{W41}
     M^2_3 &=& M^2_4 + 6\,(B_3)^2 \,,\\
\label{W42}
     M^3_3 &=& M^3_4 - 2B_3M^2_4 - 20\,(B_3)^3\,.
\end{eqnarray}
This we substitute in the expression for $M^4_4$ to obtain:
\begin{eqnarray}
\label{W43}
  M^4_4&=&3\left\{B_3M^3_4 -3\,(B_3)^2M^2_4 -27\,(B_3)^4\right\}\,.
\end{eqnarray}
The new spin-4 generator $w^4_4$ is then defined as
 the linear combination
\begin{eqnarray}
\label{W44}
  w^4_4 &=&M^4_4-3\left\{B_3M^3_4 -3\,(B_3)^2M^2_4\right\}\nonumber\\
        &=&-81\,(B_3)^4\,.
\end{eqnarray}
Now let us define $w^3_4$. We get from Table 2:
\begin{eqnarray}
\label{W45}
  M^3_4&=&M^3_3+2B_3M^2_3+8\,(B_3)^3
\end{eqnarray}
To express $M^3_3$ in terms of $M^l_4$, $l<3$ we must first make use
 of the $N=3$ entries in Table 2. These allow us to express $M^3_3$
 in terms of $M^2_3$:
\begin{eqnarray}
\label{W46}
   M^3_3&=& 2B_2M^2_3+8(B_2)^3 \,.
\end{eqnarray}
This, and (\ref{W41}), is then substituted in (\ref{W45}).
The result is:
\begin{eqnarray}
\label{W47}
  M^3_4&=&2(B_2+B_3)(M^2_4 + 6\,(B_3)^2) +8\,(B_2)^3+8\,(B_3)^3\,.
\end{eqnarray}
 For $M^3_4$
 we now make a redefinition which gets rid of $M^2_4$.
 The resulting spin-3 generator $w^3_4$ is
\begin{eqnarray}
\label{W48}
  w^3_4 &=&8\,(B_2)^3+12B_2\,(B_3)^2+20\,(B_3)^3\,.
\end{eqnarray}
After employing a similar procedure for
 the spin-2 generator we find that there is no redefinition to be made.
 The result is:
\begin{eqnarray}
\label{W49}
  w^2_4 &=& M^2_4 = -\,(B_1)^2 -3\,(B_2)^2 - 6\,(B_3)^2\,.
\end{eqnarray}
The algorithm relies on the use of the inverse of (\ref{Miura1}).
 To complete the redefinition for $w_4$ required the inverse of
 $a_{l,k}{}^{N+1}$ for all $N<4$.

Let us now consider the above algorithm for general $N$.
 We start with the highest spin of
 the $w_{N+1}$-algebra.  From (\ref{Miura1})
 and (\ref{adef}) we obtain for this generator:
\begin{eqnarray}
\label{MN+1}
  M_{N+1}^{N+1} &=& \sum_{l=0}^{N} (-1)^{N-l} N\,(B_N)^{N+1-l}\,
           M^{l}_N  \,.
\end{eqnarray}
Now, (\ref{Miura1}) expresses the generators of $w_{N+1}$ in terms
 of those of $w_{N}$, but, as in (\ref{W41}-\ref{W42}),
 we can use (\ref{Miura1}) in the opposite direction
 to express the $M^{l}_N,\ l=0,\ldots,N$
 in terms of $M^k_{N+1},\ k=0,\ldots,N$.
 As we saw in the $w_4$-example above, this requires
 the inverse of the $(N+1)\times (N+1)$
 lower triangular matrix $a_{l,k}{}^{N+1}$,
 ${l,k}=0,\ldots N$. The inverse takes on the following form:
\begin{eqnarray}
 k\le l\ &&
   f_{l,k}{}^{N+1} = \sum_{m=0}^{l-k} \left({N-k-m\atop l-k-m}\right)
           \,(-N)^m \nonumber \\
\label{fdef}
       &&\phantom{f_{l,k}{}^{N+1}} = \left({N-k\atop l-k}\right)
            {}_2F_1(1,-l+k;-N+k;-N) \,,\\
l< k\le N\ &&
         f_{l,k}{}^{N+1} = 0 \,.
\end{eqnarray}
The inverse of (\ref{Miura1}) then becomes:
\begin{eqnarray}
\label{Miura2}
   M^l_{N} &=& \sum_{k=0}^N f_{l,k}{}^{N+1} \,(B_N)^{l-k} M^{k}_{N+1}\,.
\end{eqnarray}
This we can substitute in (\ref{MN+1}), to obtain:
\begin{eqnarray}
  M^{N+1}_{N+1} &=&
   (-1)^N N\sum_{k=0}^N\,(B_N)^{N+1-k} M^k_{N+1}\sum_{l=0}^{N}(-1)^l
           f_{l,k}{}^{N+1} \nonumber\\
\label{MN+1res}
  &=& -\sum_{k=0}^{N} \,(-NB_N)^{N+1-k} M^{k}_{N+1} \,.
\end{eqnarray}
Here we have used the following result for the coefficients $f$:
\begin{eqnarray}
 \sum_{l=0}^{N}(-1)^l
           f_{l,k}{}^{N+1} &=& (-1)^k N^{N-k} \,.
\end{eqnarray}
Now we can redefine the highest spin (we will denote the
 spin $l$ generator of $w_N$
 in the new basis by $w_N^l$):
\begin{eqnarray}
\label{eq:highest}
  w^{N+1}_{N+1} &=& M_{N+1}^{N+1}+\sum_{k=2}^{N}
    \,(-NB_N)^{N+1-k} M^{k}_{N+1}   \nonumber \\
\label{wN+1}
    &=& (-1)^N \,(N B_N)^{N+1}\,.
\end{eqnarray}
Note that we only use $M^k_{N+1}$ for $k=2,\ldots,N$ in the redefinition,
 since $M^0_{N+1}$ and $M^1_{N+1}$ are field independent
 constants (\ref{M0}-\ref{M1}),
 which are not generators of the $w_{N+1}$-algebra.

To obtain $w^{N}_{N+1}$ we start with
\begin{eqnarray}
\label{MN}
  M^{N}_{N+1} &=& M^N_N +\sum_{l=0}^{N-1} a_{N,l}{}^{N+1}
     M^{l}_{N} \,(B_N)^{N-l}  \,.
\end{eqnarray}
We then rewrite $M^N_N$ using our result (\ref{MN+1res})
 with $N+1\to N$. In the second term of (\ref{MN}) we substitute
 (\ref{Miura2}). The result is
\begin{eqnarray}
  M^N_{N+1} &=& -\sum_{k=0}^{N-1} \sum_{l=0}^{N-1}
   M^k_{N+1} \{-(N-1)B_{N-1}\}^{N-l} (B_N)^{l-k} f_{l,k}{}^{N+1}
   \nonumber \\
\label{MN1}
 && + \sum_{k=0}^{N-1} M^{k}_{N+1} (B_N)^{N-k}
      \sum_{l=0}^{N-1} a_{N,l}{}^{N+1}f_{l,k}{}^{N+1}
\end{eqnarray}
The last sum can be rewritten using
\begin{eqnarray}
  \sum_{l=0}^{N-1} a_{N,l}{}^{N+1}f_{l,k}{}^{N+1}
    &=& \delta_{N,k}-a_{N,N}{}^{N+1}f_{N,k}{}^{N+1}
    \nonumber\\
  &=& \delta_{N,k}-f_{N,k}{}^{N+1}
\end{eqnarray}
Substituting this back in (\ref{MN1}) gives finally:
\begin{eqnarray}
\label{MNres}
   M^N_{N+1} &=& -\sum_{k=0}^{N-1} \sum_{l=0}^{N}
   M^k_{N+1} \{-(N-1)B_{N-1}\}^{N-l} (B_N)^{l-k} f_{l,k}{}^{N+1} \,.
\end{eqnarray}
Again we can redefine to obtain the generator $w^N_{N+1}$:
\begin{eqnarray}
\label{wN}
     w^N_{N+1} &=& -\sum_{l=0}^{N}
    \{-(N-1)B_{N-1}\}^{N-l} (B_N)^{l} f_{l,0}{}^{N+1} \,.
\end{eqnarray}
Note that the $l=0$ term, which is independent of $B_N$, is equal to
 $w^N_N$.

This procedure can be continued for all spins.
 To continue to lower spins one needs to determine
 for each $l$ the analogue of (\ref{MN+1res}, \ref{MNres}), since,
 as for $l=N$, one uses the result for spin $l+1$ in the calculation
 for spin $l$. The redefinition then amounts to throwing away all
 contributions of $M^l_{N+1}$ in the result except that of $l=0$.
 For spin $l=N-1$ we obtain in this way:
\begin{eqnarray}
\label{wN-1}
   w^{N-1}_{N+1} &=& -\sum_{k,l=0}^{N-1}
  \{-(N-2)B_{N-2}\}^{N-1-l}(B_{N-1})^{l-k} (B_N)^{k}
    f_{l,k}{}^{N} f_{k,0}{}^{N+1} \,,
\end{eqnarray}
from which one can generalise to arbitrary spins:
\vspace{.1truecm}
\begin{eqnarray}
\label{eq:formula}
  w^{N-l}_{N+1} &=& -\sum_{k_1,\ldots,k_{l+1}=0}^{N-l}
  \{-(N-l-1)B_{N-l-1}\}^{N-l-k_1} \times{}
 \nonumber\\
  &&\quad{}\times (B_{N-l})^{k_1-k_2} \cdots
     (B_{N-1})^{k_{l}-k_{l+1}} (B_N)^{k_{l+1}} \times{}
  \nonumber\\
\label{wN-n}
  &&{}\times f_{k_1,k_2}{}^{N-l+1}f_{k_2,k_3}{}^{N-l+2}\cdots
      f_{k_{l},k_{l+1}}{}^{N}f_{k_{l+1},0}{}^{N+1} \,,
\end{eqnarray}
 for $l=0,\ldots,N-2$. The highest-spin generator $w^{N+1}_{N+1}$
 is given in (\ref{wN+1}).
 Again, if we select the term with vanishing power of $B_N$, we
 obtain $w^{N-l}_{N}$. It is a simple exercise to show that for
 $l=N-2$ the generator $w^2_N$ is equal to (\ref{M2}), i.e., the
 energy-momentum tensor is not modified by our redefinitions.

So in our new basis
 we have obtained in (\ref{eq:highest}), (\ref{eq:formula}) closed formulae
 for all generators of the
 classical $w_{N}$-algebra. Closure is
 guaranteed because of the closure of the algebra in the Miura basis.
 Of course, it is a formidable exercise to obtain the structure constants
 and the corresponding classical BRST-charge explicitly for
 the complete $w_N$-algebra. In the next section, where we will address
 these problems, we will therefore limit ourselves to the
 $v^{N}_{N+1}$-algebra, which consists of the generators $w^N_{N+1}$
 and $w^{N+1}_{N+1}$.

For future reference we give explicit results for the
 redefined generators for the algebras $w_N,\ N=2,3,4,5$ in Table 3.

\begin{center}
\begin{tabular}{|l||l|}
\hline
     &\\
$N=2$&$w^2_2=-\,(B_1)^2$\\
     &\\
$N=3$&$w^2_3=w^2_2-3\,(B_2)^2$\\
     &$w^3_3=8\,(B_2)^3$\\
     &\\
$N=4$&$w^2_4=w^2_3-6\,(B_3)^2$\\
     &$w^3_4=w^3_3+12B_2(B_3)^2+20\,(B_3)^3$\\
     &$w^4_4=-81\,(B_3)^4$\\
     &\\
$N=5$&$w^2_5=w^2_4-10\,(B_4)^2$\\
     &$w^3_5=w^3_4+20B_2\,(B_4)^2+20B_3\,(B_4)^2+40\,(B_4)^3$\\
     &$w^4_5=w^4_4-90\,(B_3)^2\,(B_4)^2-120B_3\,(B_4)^3-205\,(B_4)^4$\\
     &$w^5_5=1024\,(B_4)^5$\\
     &\\
\hline
\end{tabular}
\end{center}

\vspace{.5truecm}

\noindent {\bf Table\ 3.} \ \ \ \ \ Generators of $w_N$ in our
new basis for some low values of N.

\section{The $v^{N+1}_{N+1}$ and $v^{N}_{N+1}$ subalgebras}

The advantage of the new basis introduced in the previous section
 is that for each subalgebra of $w_N$ one can define a nilpotent
 BRST-charge $Q$. Clearly the $Q$'s, as the subalgebras, form a
 nested structure, in which $Q^s_N$, the BRST-charge for
 the $v^{s}_N$-subalgebra, contains as contributions all $Q^{s'}_N$
 for $s'\ge s$. Since each of these $Q$'s is separately nilpotent,
 this nested structure should simplify the construction of, e.g.,
 the physical states of the corresponding quantum theory, assuming
 of course that a quantum extension of this nested structure can
 be given. In this and the next
 section we will further discuss the classical
 structure of the algebra and its BRST-current. The quantum
 extension will be considered in some specific examples in
 Section 5.

For simplicity, let us start with the $v^{N+1}_{N+1}$-algebra. Its
 only generator is given in (\ref{wN+1}). It is a simple matter
 to calculate the Poisson bracket with itself.
The basic OPE is given by\footnote{
In order to facilitate the transition to the quantum case, it is convenient
to represent the Poisson brackets by
 Operator Product Expansions, in which only {\sl single} contractions of
 fields are considered. After quantisation {\sl multiple} contractions have to
be taken into account as well.}
\begin{eqnarray}
\label{BOPE}
   B_m(z) B_n(w) &\sim& {\delta_{mn}\over 2n(n+1)} {1\over (z-w)^2}
         \,.
\end{eqnarray}
For the generator $w^{N+1}_{N+1}$ we then find:
\begin{eqnarray}
   w^{N+1}_{N+1}(z)w^{N+1}_{N+1}(w)&\sim&
    {\textstyle{1\over 2}}(-1)^N N^N\,(N+1)\times
       \nonumber\\
\label{OPEN+1}
 && \times \bigg[{1\over (z-w)^2} + {\textstyle{1\over 2}}
   {\partial\over z-w}\bigg]\bigg[
         (B_N)^{N-1}w^{N+1}_{N+1}(w)
              \bigg]\,.
\end{eqnarray}
The BRST-current for the algebra (\ref{OPEN+1}) is easily obtained.
 Introducing the ghost and antighost pair $(c_{N+1},\ b_{N+1})$,
 with the contraction
\begin{eqnarray}
    c_l(z)b_k(w) &\sim& {\delta_{lk}\over z-w}
\end{eqnarray}
 for any $l,k$, we obtain
\begin{eqnarray}
\label{eq:jBRST}
   j^{N+1}_{N+1} = c_{N+1}w^{N+1}_{N+1}
     -{\textstyle{1\over 4}}(-1)^N N^N\,(N+1)(B_N)^{N-1}
       \partial c_{N+1}c_{N+1}b_{N+1} \,.
\end{eqnarray}
The pole of order one in the OPE of $j$ with itself is a total
 derivative:
\begin{eqnarray}
   j(z) j(w) &\sim& \ldots + {\partial(\ldots)\over z-w} + \ldots \,,
\end{eqnarray}
so that $Q=\oint\,dz j(z)$ satisfies $\{Q,Q\}=0$.

Thus we see that the BRST-current for the $v^{N+1}_{N+1}$-algebra
 contains terms that are no more than cubic in the ghosts. This
 feature is no longer present  when we consider
 the $v^{s}_{N+1}$-algebra.

For general $N$ we will only consider
 the algebra containing the two generators (\ref{wN+1})
 and (\ref{wN}).  In this case we can obtain the structure
 constants of the algebra explicitly in terms
 of the coefficients $f$ as given in (\ref{fdef}).
 The  $v^{N}_{N+1}$-algebra is given by the
 OPE's (\ref{OPEN+1}) and the following ones
\begin{eqnarray}
  && w_{N+1}^{N+1}(z)w_{N+1}^{N}(w) \sim \nonumber\\
  &&\quad
    -{\textstyle{1\over 2}}
      \sum_{k=1}^{N}(-1)^{N-k} k f_{k,0}{}^{N+1}
        ((N-1)B_{N-1}(w))^{N-k}(B_N(w))^{k-2} \times {}
     \nonumber\\
  \label{OPENN+1}
   &&\qquad\times\bigg[
      { w^{N+1}_{N+1}(w) \over N(z-w)^2} +
      {\partial w^{N+1}_{N+1}(w) \over (N+1)(z-w)}\bigg] \,,\\
   && w_{N+1}^{N}(z)w_{N+1}^{N}(w) \sim \nonumber\\
 &&\quad
  \bigg[{1\over (z-w)^2} + {\textstyle{1\over 2}}
     {\partial\over z-w}\bigg] \times \nonumber\\
 &&\qquad\times \bigg[{\textstyle{1\over 2}}\sum_{k=0}^{N-2}
    (-1)^{N+k+1}(N-k)(N-k-1)f_{k,0}{}^{N+1}\times\nonumber\\
 &&\qquad\qquad
      \times ((N-1)B_{N-1}(w))^{N-k-2}(B_N(w))^{k}w_{N+1}^{N}(w)
    \nonumber\\
 &&\qquad + {1\over 2N}\sum_{k=0}^{N-3} (-1)^{N-k} (k+2)(N-2-k)
    f_{k+2,0}{}^{N+1}\times \nonumber\\
  \label{OPENN}
 &&\qquad\qquad\times
    ((N-1)B_{N-1}(w))^{N-k-3}(B_N(w))^{k}w_{N+1}^{N+1}(w)\bigg] \,.
\end{eqnarray}

Since the above algebra has been obtained from the Miura basis by a
 redefinition, closure is guaranteed. Nevertheless, it
 is interesting to check how restrictive the requirements of closure are
 on the coefficients in (\ref{OPEN+1}), (\ref{OPENN+1}) and
 (\ref{OPENN}). It is clear that in (\ref{OPEN+1}) there are no
 restrictions at all: for a single scalar we can always form an algebra
 with a single generator, for any spin.
 In (\ref{OPENN+1}) the
 sums must be such that negative powers of $B_N$ are avoided. This is
 indeed the case, since $f_{1,0}=0$. One can easily check that this is
 the only condition on the coefficients $f$ required for closure.
 In (\ref{OPENN}) the situation is more complicated. One can
 parametrize the right hand side of (\ref{OPENN}) with an expansion in
 powers of $B_{N-1}$ and $B_N$ with arbitrary coefficients,
 multiplying the generators $w^{N}_{N+1}$
 and $w^{N+1}_{N+1}$. It turns out that
 the requirements of closure can be solved for all unknown coefficients,
 but that two consistency equations remain. In terms of the coefficients $f$
 these are two quadratic identities of the form:
\begin{eqnarray}
  &&{(N-1)\over N} \sum_{k=0}^{N-1}(k+1)(N-k)f_{k,0}{}^{N+1}
     f_{N-1-k,0}{}^{N+1}
   \nonumber\\
  &&\quad + {1\over N(N+1)}
     \sum_{k=0}^{N-1}(k+1)(N-k)f_{N-k,0}{}^{N+1}f_{k+1,0}{}^{N+1}
   \nonumber\\
\label{ident1}
  &&\quad\quad
    -\sum_{k=0}^{N-2}
    (N-k)(N-k-1)f_{k,0}{}^{N+1}f_{N-k-1,0}{}^{N+1} = 0\,,
   \\
  &&{(N-1)\over N}\sum_{k=0}^N k(N-k)f_{k,0}{}^{N+1}f_{N-k,0}{}^{N+1}
   \nonumber\\
  &&\quad
   +{1\over N(N+1)}\sum_{k=0}^{N-2} (k+2)(N-k) f_{k+2,0}{}^{N+1}
          f_{N-k,0}{}^{N+1}
   \nonumber\\
 \label{ident2}
  &&\quad\quad
   + \sum_{k=0}^{N-2} (N-k)(N-k-1)f_{k,0}{}^{N+1}f_{N-k,0}{}^{N+1}=0\,.
\end{eqnarray}
Calculations for the
coefficients $f_{k,0}{}^{N+1}$
for general $N$ and $k=0,1,\ldots$ are done using the explicit form
(\ref{fdef}). In some calculations, such as in the check of
(\ref{ident1}, \ref{ident2}) we also need for general $N$
the coefficients $f_{N-k,0}{}^{N+1}$
for $k=0,1,\ldots$. We have then used the following representation
of the $f$'s:
\begin{eqnarray}
   f_{N-k,0}{}^{N+1} &=& \sum_{l=0}^k \left({N-l\atop k-l}\right)
    {(N)^k\over (1+N)^{k+1}}\times
  \nonumber\\
 &&\quad
  \times\bigg[ 1-(-N)^{N-k+1}\sum_{j=0}^{l}
   \left({N-k+1\atop j}\right) \left(-\, {1+N\over N}\right)^j
        \bigg] \,.
\end{eqnarray}

The BRST-charge
 for the $v^N_{N+1}$-algebra is much more complicated
 than (\ref{eq:jBRST}) for the $v^{N+1}_{N+1}$-algebra.
In particular,
 there will be ghost contributions of higher order than cubic terms.
The same applies
to the BRST-charge for the general $v^l_{N+1}$-algebra.
 We have not attempted to obtain the BRST-current $j_{N+1}^l$
 for general $l$ and N. Instead, we will give in the next section explicit
expressions for some specific values of $l$ and $N$.

Using a dimensional
 argument, it is possible to give a limit on the terms of higher-order
 in the ghost fields that may appear in the BRST-charge.
 Let us
 briefly present this argument for the $v^N_{N+1}$-algebra. There we
 have two pairs of ghosts, $b_{N+1},\ c_{N+1}$ and $b_N,\ c_N$.
 The conformal spin of the BRST current equals 1, the ghost fields $b_n$
 and $c_n$ have spins $n$ and $1-n$. Also, $Q$ has ghost number 1.
 A $2n+1$-order ghost contribution to $Q$ for the $v^N_{N+1}$-algebra
 would be of the form:
\begin{eqnarray}
\label{ghostprod}
  && (b_{N+1})^k (b_N)^l (c_{N+1})^p (c_N)^q \,,
   \quad k+l=n,\ p+q=n+1 \,,
\end{eqnarray}
where the powers of the anti-commuting ghost fields are given by, e.g.,
 $(b_n)^k\equiv b_n (\partial b_n) \ldots (\partial^{k-1} b_n)$.
 The conformal weights $s_b$ and $s_c$ of the product of all $b$-
 and $c$-ghosts in (\ref{ghostprod}) is then
\begin{eqnarray}
   s_b &=& k^2 +k(1-n)+nN +{\textstyle{1\over 2}}n(n-1) \,,
  \nonumber\\
   s_c &=& p^2-2p-np-(n+1)N+{\textstyle{1\over 2}}(n+1)(n+2) \,.
  \nonumber
\end{eqnarray}
 The minimum values for $s_b$ and $s_c$ are reached for $k=(n-1)/2$
 and $p=(n+2)/2$, respectively. The value of the sum of the minima
 of $s_b$ and $s_c$ is given by:
\begin{eqnarray}
\label{smin}
   s_{min} &=& {\textstyle{1\over 4}}(2n^2+2n-1)-N \,.
\end{eqnarray}
For such a ghost term to exist in the BRST-current we must have
 $s_{min}\le 1$, so that it is possible
 to obtain $s_Q=1$.
 Therefore we should have
\begin{eqnarray}
\label{ghostlimit}
    2n^2+2n-1 \le 4(N+1)
\end{eqnarray}
for the $v^N_{N+1}$-algebra. For the $v^3_4$-algebra this implies that
 terms of fifth order in the ghosts can be written down. However, as
 we shall see in the next section, only cubic ghost-terms actually
 appear. For the $v^4_5$-algebra fifth order ghost terms are possible, but
 seventh order ghost terms are not. In that case we find that the
 fifth order terms are required in the BRST-charge.

Clearly, the dimensional argument can
 be extended to $v^l_{N+1}$-algebras.

\section{Classical BRST charges}

\def\da{\partial\phi_1}
\def\db{\partial\phi_2}
\def\dc{\partial\phi_3}

In this section we give explicit expressions for the
BRST charges of $w_3, w_4$ and the subalgebra $v_5^4 \subset w_5$
in the new basis,
thereby making explicit the nested structure (\ref{eq:Qnest}).
To obtain these BRST charges, it is convenient to use an
iterative procedure. Starting from the terms in the
BRST charge that are linear in the ghosts (the terms
containing the generators), one obtains higher order ghost
terms by demanding nilpotency. In the next order one finds
that
the coefficients multiplying the cubic ghost terms are
the structure constants of the algebra (in the new basis). Since we are
dealing with field dependent structure constants, it may be necessary
to add higher order ghost terms as well.

For pedagogical reasons, we will discuss first the case
of the $w_3$ algebra in somewhat more detail \cite{po1}. The generators of the
$w_3$-algebra are given in Table 3. Using (\ref{Bdef}),
these generators can be written as
\begin{eqnarray}
T &=& -2\, w_3^2\, =\, -{1\over 2}(\da)^2 - {1\over 2}(\db)^2 \,,
\nonumber\\
\label{w3N}
W &=& -2\sqrt{3}\, w_3^3\, =\, {2i\over 3}(\db)^3 \,.
\end{eqnarray}
Note that the generators $w_3^2$ and $w_3^3$ have been rescaled.
This makes $T$ an energy-momentum tensor generating the
Virasoro algebra. For $W$ the rescaling is just a matter of
convenience.

The OPE of $W$ with itself is\footnote{The OPE's involving the
energy-momentum tensor $T$ are standard and not given here.}
\begin{eqnarray}
W(z)W(w)\, \sim\, \bigg[{1\over (z-w)^2}
  + {\textstyle{1\over 2}}{\partial\over z-w}\bigg](-6i\db W) \,.
\end{eqnarray}
{}From this algebra one can read of the BRST current $j_3(z)$
up to third order ghost terms, and it turns out that no higher order
terms are needed. It can be written as $j_3(z)=j_3^2(z)$ with
\begin{eqnarray}
\label{eq:jw3}
j_3^3 (z)&=&c_3 W - 3i\db c_3 \partial c_3 b_3 \,.\nonumber\\
j_3^2(z)&=&
c_2 (T+T_{c_3,b_3}+{1\over2}T_{c_2,b_2}) + j^3_3 (z)\,,
\end{eqnarray}
where we defined the ghost energy-momentum tensors
\begin{eqnarray}
\label{ghT}
T_{c_s,b_s} &=& -s b_s \partial c_s - (s-1)\partial b_s c_s
\end{eqnarray}
for arbitrary spin $s$.
The expression for $j_3^3(z)$ agrees with the formula for
general $N$ given in (\ref{eq:jBRST}).
Note that the two charges $Q_3^3$ and $Q_3^2-Q_3^3$ are separately nilpotent.

It is instructive to compare the above result for the BRST charge in the
new basis with the
one in the Miura basis. The two expressions are related to each other
by a canonical transformation in the extended phase space
\cite{Henn}. It turns out that the canonical transformation that relates
(\ref{eq:jw3}) to the Miura basis is generated by
\begin{eqnarray}
G=i\db c_3 b_2\,.
\end{eqnarray}
The exponential action of the generator $G$ on an
extended phase space function $F$ is, in OPE language
\begin{eqnarray}
\label{cantr}
F(w) \rightarrow F(w) + \oint {dz\over 2\pi i}G(z)F(w)+
{1\over 2!}\oint {dz\over 2\pi i}G(z) \oint {dx\over 2\pi i}G(x)F(w)
+...\,\,.
\end{eqnarray}
This results in the following transformations of the
basic fields \cite{po1}:
\begin{eqnarray}
{\tilde{c}}_2 &=& c_2 +i\db c_3 +{1\over2}c_3 \partial c_3 b_2\,,
\nonumber\\
{\tilde{b}}_2 &=& b_2\,,\nonumber\\
{\tilde{c}}_3 &=& c_3\,,\nonumber\\
{\tilde{b}}_3 &=& b_3 -i\db b_2 +{1\over2}b_2 \partial b_2 c_3\,,
\nonumber\\
\label{tr}
\partial{\tilde{\phi}}_2 &=& \db +i\partial (b_2 c_3)\,,
\end{eqnarray}
where the tilde indicates the fields in the Miura basis.
Due to anticommutativity of the ghost variables,
only the first few terms in (\ref{cantr}) contribute to
(\ref{tr}).
The BRST current (\ref{eq:jw3}) now transforms into its
Miura form (suppressing the tilde on both fields and generators) \cite{thm1}
\begin{eqnarray}
\label{eq:jw3m}
j(z)=c_2(T + T_{c_3,b_3}+
{1\over2}T_{c_2,b_2}) + c_3 W
+{1\over2}c_3 \partial c_3 b_2 T \,.
\end{eqnarray}
Note that the nested structure is absent in the Miura basis: the
BRST current (\ref{eq:jw3m}) cannot be written as the sum of two
separate nilpotent currents.

The advantage of using the new basis instead of the Miura basis
is even more apparent when we discuss $w_4$. The BRST charge for $w_4$
in the Miura basis has recently been calculated in \cite{horn,zhu2}.
The authors of \cite{horn,zhu2} find that the BRST current contains
terms up to seventh order in the
ghosts. As we will show below, in the new basis we not only make
the nested structure of the BRST charge explicit, but we furthermore find
that in the new basis all higher order ghost terms vanish and that at most
trilinear ghost terms occur.

The generators of the $w_4$-algebra in the new basis are given
in Table 3:
\begin{eqnarray}
T &=& -2\, w_4^2 \,=\, -{1\over2}(\da)^2 -{1\over2}(\db)^2
-{1\over2}(\dc)^2\,,\nonumber\\
W &=& 3i\sqrt{3}\, w_4^3 \,=\,(\db)^3 +{3\over4}\db(\dc)^2
+{5\sqrt{2}\over8}(\dc)^3\,,\nonumber\\
V &=& -{64\over9}\,w_4^4 \,=\,(\dc)^4\,.
\end{eqnarray}
where we have made some convenient rescalings of the generators.
The nontrivial OPE's (not involving the energy momentum tensor)
among these generators are
\begin{eqnarray}
W(z)W(w) &\sim &
 \bigg[{1\over (z-w)^2} + {\textstyle{1\over 2}}{\partial\over z-w}\bigg]
 (-9\db W-{243\over32}V) \,,\nonumber\\
W(z)V(w) &\sim & {-6\db V -9\dc V\over(z-w)^2}+
{-{3\over2}\db \partial V-6\partial^2\phi_2 V-{18\over5}\partial (\dc V)
\over(z-w)}\,,\nonumber\\
\label{w4N}
V(z)V(w) &\sim &
\bigg[{1\over (z-w)^2} + {\textstyle{1\over 2}}{\partial\over z-w}\bigg]
  \bigl (-16(\dc)^2 V\bigr )                  \,.
\end{eqnarray}
{}From this algebra, one can read off the BRST current $j_4(z)$ up to
third order ghost terms, and it turns out that no higher order
terms are needed. It can be
written as $j_4(z)=j^2_4(z)$ with
\begin{eqnarray}
j_4^4(z) &=& c_4 V -8(\dc)^2 c_4\partial c_4 b_4\,,
 \nonumber \\
j^3_4(z) &=& c_3 W-{9\over2}\db c_3\partial c_3 b_3-{243\over64}
c_3 \partial c_3 b_4 -{9\over2}\db c_3 c_4 \partial b_4
\nonumber\\ & &-6\db c_3\partial c_4 b_4
+{9{\sqrt{2}}\over2}\dc\partial c_3 c_4 b_4 -3\sqrt{2}\dc c_3 c_4
\partial b_4 \nonumber\\
 && + j_4^4(z) \,,\nonumber\\
j^2_4(z) &=& c_2(T+T_{c_3,b_3}+T_{c_4,b_4}+{1\over2}T_{c_2,b_2})
   + j^3_4(z) \,.
\end{eqnarray}
The nested structure of the BRST charges manifests itself through
the fact that the BRST charges associated with
$j_4^4\,,j^3_4$ and $j^2_4$ are all nilpotent. Furthermore, $j^2_4-j^3_4$
is the BRST-current of the Virasoro algebra, and is separately
nilpotent.

Sofar, for $w_3$ and $w_4$ in the new basis, we have not encountered terms
in the BRST current that are of
higher than third order in the ghosts. This is not always the case.
The most simple example where one has to go beyond the trilinear ghost
terms, even in the new basis, is given by the BRST current corresponding to
the $v_5^4$ subalgebra of $w_5$. In particular, we find that the
BRST current $j_5^4$ contains terms quintic in the ghosts. The nested
structure of the currents is given by:
\begin{eqnarray}
j_5^5 &=& c_5\{{4i\over 125}\sqrt{10}(\partial\phi_4)^5\}
+{2i\over 5}\sqrt{10}c_5'c_5b_5(\partial\phi_4)^3\nonumber\\
j_5^4 &=& c_4\{-{9\over 64}(\partial\phi_3)^4
-{3\over 32}(\partial\phi_3)^2(\partial\phi_4)^2
-{\sqrt{15}\over 40}\partial\phi_3(\partial\phi_4)^3
-{41\over 320}(\partial\phi_4)^4\}\nonumber\\
&+&\{-{9\over 8}(\partial\phi_3)^2-{1\over 8}(\partial
\phi_4)^2\}c_4'c_4b_4\nonumber\\
&+&\{{-3i\over 8}\sqrt{10}\partial\phi_4-{5i\over 8}\sqrt{6}\partial\phi_3
\}c_4'c_4b_5\\
&+&\{-{15\over 16}(\partial\phi_3)^2-{\sqrt{15}\over 8}\partial\phi_3
\partial\phi_4-{19\over 16}(\partial\phi_4)^2\}c_5'c_4b_5\nonumber\\
&+&\{{\sqrt{15}\over 4}\partial\phi_3\partial\phi_4
+{11\over 8}(\partial\phi_4)^2\}\nonumber\\
&+&\{-{3\over 4}(\partial\phi_3)^2-{1\over 8}(\partial\phi_4)^2\}
c_5c_4b_5'\nonumber\\
&+&{\sqrt{15}\over 4}\partial^2\phi_3\partial\phi_4c_5c_4b_5\nonumber\\
&-&{5\over 4}c_5'c_5c_4b_5'b_5 +{5\over 4}c_5'c_4'c_4b_5b_4
+c_5c_4'c_4b_5'b_4 \nonumber\\
&+& j^5_5 \,. \nonumber
\end{eqnarray}

\section{Quantisation}

Sofar, our discussion has basically been at the classical level.
In this section we will discuss some aspects of the quantisation, in particular
the construction of the quantum BRST operators. The results of
this section indicate that the nested structure found at the
classical level survives the quantisation.

Our strategy is to use the classical results of the
previous sections as a starting point for the construction of the quantum
BRST operators\footnote{Note that we write the quantum
expressions with boldface.}.
In practice, the easiest way to obtain explict expressions for
the BRST operators for low values of $N$ is to parametrize all possible
quantum corrections to the classical BRST charge, and then to determine the
coefficients occurring in the Ansatz by requiring nilpotency of the
quantum BRST operator. We will use this explicit method to discuss the
quantisation of the $w_4$ algebra\footnote{The quantisation of the
$w_3$-algebra in the new basis was done in \cite{po1} and we will not
repeat it here.}.

We would like to stress that the use of the new basis greatly
 facilitates the construction of the quantum BRST operator.
The nested structure enables one to construct the BRST operator
in an iterative way. One starts with
 ${\bf Q}^N_N$, the BRST-operator corresponding to the highest spin generator
 of $W_N$. This will depend on only one scalar, and on the spin-N ghosts
 $b_N,\ c_N$. Next one goes on to ${\bf Q}^{N-1}_N$, which will depend
 on one additional scalar and the spin-$(N-1)$ ghost pair as well. In this
 way, one obtains at each level a nilpotent BRST-operator, which contains
 the operators of the higher-spin subalgebras.
In the last step one obtains the BRST operator
of the complete $W_N$ algebra.

For $N=4$ the quantum extension ${\bf j}_4^4$ of the highest-spin contribution
$j^4_4$ to the classical BRST current was already given in
\cite{po3}. We now give the result for the full $w_4$-algebra, including
 also ${\bf j}^3_4$ and  ${\bf j}^2_4$:
\begin{eqnarray}
{\bf j}_4^4 &=& c_4\{(\partial\phi_3)^4 + {18\over 5}\sqrt{15}\partial^2\phi_3
(\partial\phi_3)^2 + {41\over 5}\partial^2\phi_3\partial^2\phi_3\nonumber\\
&+& {124\over 15}\partial^3\phi_3\partial\phi_3
+{23\over 75}\sqrt{15}\partial^4\phi_3\}\nonumber\\
\label{j41}
&-&8(\partial\phi_3)^2c_4c_4'b_4 + {8\over 5}\sqrt{15}\partial^2\phi_3
c_4c_4'b_4 \\
&+&{16\over 5}\sqrt{15}\partial\phi_3c_4c_4''b_4
+{4\over 5}c_4c_4'''b_4 - {16\over 3}c_4c_4'b_4''\,,\nonumber\\
{\bf j}_4^3 &=&
c_3\{(\partial\phi_2)^3 + {3\over 4}\partial\phi_2(\partial\phi_3)^2 +
{5\over 8}\sqrt{2}(\partial\phi_3)^3\nonumber\\
&+&{27\over 20}\sqrt{30}\partial\phi_2\partial^2\phi_2 + {27\over20}\sqrt{15}
\partial\phi_2\partial^2\phi_3 + {81\over 40}\sqrt{30}
\partial\phi_3\partial^2\phi_3\nonumber\\
\label{j42}
&+& {93\over 40}\partial^3\phi_2 + {69\over 10}\sqrt{2}\partial^3\phi_3\}\\
&-& {9\over 2}\partial\phi_2c_3c_3'b_3 - {27\over 40}\sqrt{30}c_3''c_3b_3
- {243\over 64}c_3c_3'b_4\nonumber\\
&-&{9\over 2}\partial\phi_2c_3c_4b_4' - 6\partial\phi_2c_3c_4'b_4
-{81\over 40}\sqrt{30}c_3''c_4b_4\nonumber\\
&+&{27\over 40}\sqrt{30}c_3c_4''b_4 + {9\over 2}\sqrt{2}\partial\phi_3
c_3'c_4b_4 - 3\sqrt{2}\partial\phi_3c_3c_4'b_4\nonumber\\
\label{j43}
 &+& {\bf j}^4_4 \,, \\
{\bf j}_4^2 &=&
 c_2 \{-{1\over 2}(\partial\phi_1)^2 -{1\over 2} (\partial\phi_2)^2
-{1\over 2}(\partial\phi_3)^2 \nonumber\\
&\pm& {9\over 20}\sqrt{10}\partial^2\phi_1
-{9\over 20}\sqrt{30}\partial^2\phi_2 - {9\over 10}\sqrt{15}\partial^2\phi_3\}
\nonumber\\
&+&c_2c_2'b_2 +3c_2c_3'b_3+2c_2c_3b_3'+4c_2c_4'b_4 + 3c_2c_4b_4'
   \nonumber\\
\label{j44}
&+& {\bf j}^3_4 \,.
\end{eqnarray}
It turns out that there exists another nilpotent BRST charge
for the quantum $W_4$-algebra
which has a different sign
for the background charge of $\phi_2$. So ${\bf j}_4^2-{\bf j}_4^3$ is the
same except that
\begin{equation}
-{9\over 20}\sqrt{30}\partial^2\phi_2 \rightarrow +{9\over 20}
     \sqrt{30}\partial^2\phi_2
\end{equation}
Using the other choice of sign for the background charge, we find
that ${\bf j}_4^4$ is the same but that ${\bf j}_4^3$ is now given by
\begin{eqnarray}
{\bf j}_4^3 &=&
c_3\{(\partial\phi_2)^3 + {3\over 4}\partial\phi_2(\partial\phi_3)^2 +
{5\over 8}\sqrt{2}(\partial\phi_3)^3\nonumber\\
&-&{27\over 20}\sqrt{30}\partial\phi_2\partial^2\phi_2 + {27\over20}\sqrt{15}
\partial\phi_2\partial^2\phi_3 + {27\over 20}\sqrt{30}
\partial\phi_3\partial^2\phi_3\nonumber\\
&+& {93\over 40}\partial^3\phi_2 - {177\over 80}\sqrt{2}\partial^3\phi_3\}\\
&-& {9\over 2}\partial\phi_2c_3c_3'b_3 + {27\over 40}\sqrt{30}c_3''c_3b_3
- {243\over 64}c_3c_3'b_4\nonumber\\
&-&{9\over 2}\partial\phi_2c_3c_4b_4' - 6\partial\phi_2c_3c_4'b_4
+{27\over 40}\sqrt{30}c_3c_4''b_4 \nonumber\\
&-&{27\over 40}\sqrt{30}c_3'c_4'b_4 + {9\over 2}\sqrt{2}\partial\phi_3
c_3'c_4b_4 - 3\sqrt{2}\partial\phi_3c_3c_4'b_4\nonumber\\
&+&{\bf j}^4_4 \,.
\end{eqnarray}
It is not clear to us whether this second solution can be related to
the first one by a canonical transformation.

Our result for the $W_4$-algebra is based on one of
 the solutions for ${\bf j}_4^4$ obtained in \cite{po3},
 namely the solution where the background
 charge of the fields are the same as in the Miura basis.
Besides this solution, the authors of \cite{po3} found
one additional
 solution for ${\bf j}^4_4$ with a different value of the background charge
 for $\phi_3$.
 We have attempted to extend also this solution with a
 ${\bf j}^3_4$ and ${\bf j}^2_4$. However, the calculation shows that for this
 additional solution such an extension is impossible.

The result (\ref{j41}) for ${\bf j}_4^4$
provides a nice example of a phenomenon which
we discussed in the introduction, namely that at the quantum level
consistency of the theory requires the existence of a nilpotent
BRST operator but not of a closed quantum algebra. Indeed, although
a nilpotent BRST operator ${\bf Q}_4^4$ exists, it is not possible to
find a quantum extension of the classical $v_4^4$-subalgebra in the
full Hilbert space\footnote{Note that it may be possible to obtain
closure by introducing additional generators besides the spin-4 generator
in the quantum algebra. This has been done for $W_3$ in \cite{hu1}.}.

The quantum BRST operator for the $W_4$-algebra in the Miura basis
 has recently been obtained in \cite{horn,zhu2}. Due to the complexity
 of their result it is hard to compare with our $N=4$ BRST-current
 (\ref{j41}-\ref{j44}) but we expect that the two expressions are
related through a canonical transformation.

\section{$W$-strings and Minimal Models}

As we already discussed in the introduction it has become more and
more clear that there exists a relation between the spectra
of $W$-strings and certain minimal
models \cite{ddr,rama,po1,po2,po3,we1,be1,hu1}.
In this section we will suggest
a very general relationship between $W$-strings and minimal
models by exploiting the nested structure discussed in this paper.
It would be interesting to see whether our suggestions can be
confirmed by explicit calculations of the spectra of $W$-strings.
We will first discuss the case of critical $W$-strings and then
investigate non-critical $W$-strings.

\subsection{Critical $W$-strings}

By a ``critical'' $W$-string, we mean that we work with only one copy of a
$W$-algebra. This $W$-algebra is realized in terms of so-called ``matter''
fields, the ``Liouville'' fields being
absent\footnote{The distinction between ``matter'' and ``Liouville''
fields is a little ambiguous, since in the case of $W$-algebras,
some of the ``matter'' fields {\sl must} have a background charge
and might therefore
also be called ``Liouville'' fields. We will adopt a convention where
the ``Liouville'' fields are introduced later as a separate realization
of the $W$-algebra (see below). This definition of a ``non-critical''
$W$-string is in accordance with the one used in \cite{ber1}.}.

As a warming-up exercise we first consider the BRST operator
${\bf Q}_N^N$,
 corresponding to the highest spin of the $W_N$-algebra. This operator
has already been constructed for $N \le 6$ in \cite{po3}.
The result, for general $N$,
 is that ${\bf Q}_N^N$ depends
  on
 a single scalar field $\phi_{N-1}$, and on the ghost fields $b_N,\ c_N$
of the spin-$N$ symmetries.
 It is nilpotent, and commutes with an energy-momentum
 tensor depending on the same fields, of the form:
\begin{eqnarray}
\label{TN}
   {\bf T}^N_N &=& -{\textstyle{1\over 2}} (\partial \phi_{N-1})^2
   - \alpha_{N-1} \partial^2 \phi_{N-1}
   - N b_N \partial c_N - (N-1) (\partial b_N) c_N \,.
\end{eqnarray}
$[{\bf Q}^N_N,{\bf T}^N_N]=0$ determines the background charge $\alpha_{N-1}$.
 For general $N$
\begin{eqnarray}
\label{aN-1}
    (\alpha_{N-1})^2 = {(N-1)(2N+1)^2 \over 4(N+1)}
\end{eqnarray}
should be one of the allowed values of the background charge. This
 has been verified for $N\le 6$ in \cite{po3}. The authors of \cite{po3}
find that also other values of the background charge are possible.
With the value of $\alpha_{N-1}$ as in (\ref{aN-1})
 we find that the total central charge of ${\bf T}^N_N$ is\footnote{Note
that this is exactly the value of the central charge corresponding to
a $SU(N-1)$ parafermionic theory \cite{we1}.}:
\begin{eqnarray}
\label{cNN}
   c^N_N &\equiv& 1 + 12(\alpha_{N-1})^2
       -2(6N^2-6N+1)= {2(N-2)\over N+1}\,.
\end{eqnarray}
This value corresponds to the central
 charge of a minimal model of the $W_{N-1}$-algebra. In general,
 the unitary
 minimal models of the $W_M$-algebra are characterized by  central
 charges (for any integer $q>M$)
\begin{eqnarray}
\label{cmin}
   c_{M,q} \equiv (M-1)\bigg[1-{M(M+1)\over q(q+1)}\bigg]\,,
\end{eqnarray}
so that (\ref{cNN}) corresponds to $c_{N-1,N}$. For $N=3$, (\ref{cNN})
then corresponds to the central charge of a Virasoro minimal model,
namely the $c={\textstyle{1\over 2}}$
Ising model. Mounting evidence that the cohomology of ${\bf Q}^3_3$
indeed produces the result of the $c={\textstyle{1\over 2}}$ Ising
model has been given in \cite{po1,po2,po3,we1,hu1}.
The relationship between critical $W_N$-strings and minimal models
for general $N$ was further explored in \cite{po4,po5,hu1}.
In particular, it was noted that
 in a particular realisation of $W_N$ \cite{po4}, the scalar fields
 $\phi_2,\ldots,\phi_{N-1}$, together with the ghost fields
 corresponding to the spins $3,\ldots,N$, form an energy-momentum
 tensor with central charge
\begin{eqnarray}
\label{c2}
   c^3_N = 1 - {6\over N(N+1)} \,,
\end{eqnarray}
corresponding to the $q=N$ minimal model of the Virasoro algebra.
We will now show, using
 the nested structure of the $W_N$-algebra, that it is possible to
 interpolate between $c^N_N$ and $c^3_N$.

The background charges of the $N-1$ scalar fields that realise the
 $W_N$-algebras are known in the Miura basis \cite{ddr,po4}.
 The iterative relation which determines
 the matter part of the energy-momentum tensor is in the
 quantum case:
\begin{eqnarray}
\label{T2N}
   {\bf T}_{N} = {\bf T}_{N-1} - {\textstyle{1\over 2}}
   (\partial \phi_{N-1})^2 + i x \sqrt{{(N-1)N\over 2}}
        \partial^2 \phi_{N-1} \,,
\end{eqnarray}
where $x$ is a parameter. The total
 central charge of all scalars is then
\begin{eqnarray}
\label{cmatter}
   c_m &=& \sum_{n=1}^{N-1} (1-6x^2 n(n+1)) =
    (N-1)(1-2x^2N(N+1)) \,.
\end{eqnarray}
On the other hand, the
 total central charge of the ghost fields is
 given by
\begin{eqnarray}
\label{cghost}
  c_{gh} &=& -2\sum_{n=2}^{N} (6n^2-6n+1) = -2(N-1)(2N^2+2N+1) \,.
\end{eqnarray}
Criticality therefore requires
\begin{eqnarray}
    x=i(2N+1)\sqrt{1\over 2N(N+1)} \,.
\end{eqnarray}
This determines the background charges of all scalar fields
 $\phi_n$:
\begin{eqnarray}
   \alpha_n &=& {2N+1\over 2}\sqrt{{n(n+1)\over N(N+1)}} \,.
\end{eqnarray}
This indeed gives (\ref{aN-1}) for $n=N-1$.

In Section 2 we performed a redefinition of the generators of the
 classical $w_N$-algebra, starting from the classical
 form of the Miura basis. In this
 redefinition the energy-momentum tensor was not modified. We conjecture
 that similarly, the energy-momentum tensor
 in our nested basis will have
 the same form as in the quantum Miura
 basis\footnote{This assumption has been verified for $N=3$ and $N=4$
(see section 5)
and for the highest-spin generator for $N\le 6$ \cite{po3}.
Note that the discussion
 of the highest spin generator
 given in \cite{hu1} depends on the same assumption.}.
 The background charges of all scalar fields are then known, and
 we can analyze the central charge of that part of the total
 energy-momentum tensor that corresponds to the BRST-operator ${\bf Q}^n_N$,
 and contains the matter fields $\phi_{n-1},\ldots,\phi_{N-1}$ and
 the ghost fields $b_n,\ c_n,\ldots,b_N,\ c_N$. The total central charge
 is given by
\begin{eqnarray}
\label{eq:relation}
   c^n_N &=& -2\sum_{k=n}^N (6k^2-6k+1) +
      \sum_{k=n-1}^{N-1}(1+12 (\alpha_k)^2) \nonumber\\
\label{cnN}
       &=& (n-2)\left\{1 - {n(n-1)\over N(N+1)}\right\} \,.
\end{eqnarray}
This is equal to $c_{n-1,N}$, the central charge of the
 $q=N$ minimal model of the
 $W_{n-1}$-algebra. For $n=2$ we find of course that $c^2_N=0$, because
 this case corresponds to
 the critical $W_N$-string. For $n=N$ we obtain
 (\ref{cNN}). Note that the relation (\ref{eq:relation}) between critical
$W_N$-strings and minimal models of the $W_{n-1}$-algebra was suggested
before, from a different point of view, in \cite{po4}.

To summarize, the nested structure of the $W_N$-algebra and of the
corresponding BRST operators clarifies the connection with minimal
models.

\subsection{Non-critical $W$-Strings}

The situation is different for the so-called non-critical
 $W_N$-string \cite{ber1}. In the case of the non-critical string
 we have classically two copies of a $w_N$-algebra,
 which we call $w_m$ and $w_l$, for matter and Liouville, respectively.
 Although the algebra is nonlinear, a combined algebra can nevertheless
 be formed with generators $w^k_N \equiv (w_m)^k_N +i^{k-2} (w_l)^k_N$.
 In the case $N=3$ the quantum BRST operator for
 this system was constructed in \cite{ber1,be3}. The noncritical
 $W_N$-string is characterized by the central charges of the matter and
 Liouville sectors, $c_m$ and $c_l$ respectively. To allow for
 a nilpotent BRST-operator these central charges must satisfy
(see  (\ref{cghost})):
\begin{eqnarray}
\label{cml}
    c_m+c_l &=& 2(N-1)(2N^2+2N+1) \,.
\end{eqnarray}
We can again go to the nested basis discussed in previous sections,
 but
 the required redefinitions can only be made
 for either the matter or the Liouville sector. Let us choose
 the Liouville sector\footnote{
The discussion below can be repeated for the case where a
 nested basis is chosen in the matter sector.}.
Then $c_l$ is given by (\ref{cmatter})
\begin{eqnarray}
\label{cl}
    c_l &=&  (N-1)(1-2x^2N(N+1)) \,,
\end{eqnarray}
but, in contradistinction to the situation considered in Section 5,
 (\ref{cml}) is now not sufficient to express $x$ in terms of $N$.
 Therefore, the non-critical strings of \cite{ber1} have one
 arbitrary parameter, $x$, which makes it possible to avoid the
 relation with minimal models. If we choose our nested basis for the
 Liouville sector, then we can make a nilpotent BRST-operator
 depending on
 the field $\phi_{N-1}$, one of the Liouville scalars, the
 spin $N$ ghost and anti-ghost fields and all fields of the
 matter sector. The total central charge corresponding to this case
 is
\begin{eqnarray}
   c^N_N &=& c_m + 1-6x^2N(N-1) -2(6N^2-6N+1) \nonumber\\
\label{cNN2}
         &=& (N-2)\left\{(2N-1)^2 +2N(N-1)x^2\right\} \,,
\end{eqnarray}
 the analogue of (\ref{cNN}). For general $x$ this does not
 correspond to a minimal model.

By choosing $x$ appropriately we can of course obtain a minimal model.
 In particular, we get the $q$'th
 unitary minimal model of the $W_{N-1}$-string
 by choosing
 $x$ equal to\footnote{Non-unitary minimal models can be obtained
 by choosing more generally $x^2=-2 -(Q_M)^2/2$, with
 $Q_M=\sqrt{p/q}-\sqrt{q/p}$ \cite{be1}. For comparison with
 Section 6.1 we will limit ourselves in the text
 to unitary models ($p=q+1$),
 but the results which follow can all be easily extended to
 the non-unitary case.}:
\begin{eqnarray}
\label{xval}
  x^2 &=& -2 - {1\over 2q(q+1)} \,.
\end{eqnarray}
Note that in
this case $c_m$, which can be determined from (\ref{cml},
 \ref{cl}), is equal to
\begin{eqnarray}
\label{cmmin}
  c_m &=& (N-1)\left\{1 - {N(N+1)\over q(q+1)}\right\} \,,
\end{eqnarray}
which corresponds to the $q$'th minimal model of the $W_N$-string.
The
values of $x$ given in (\ref{xval}) were also considered in
\cite{ber2,bo1}, where the cohomology of the non-critical $W_3$-string
was investigated.

Using the nested basis in the Liouville sector we get a series of
 nested BRST-operators, $Q^n_N$, depending on all matter fields,
 the scalars $\phi_{n-1},\ldots,\phi_{N-1}$ of the Liouville
 sector and the ghost and anti-ghost fields of the spin $n,\ldots,N$
 symmetries. For general $x$ the central charge of the
 corresponding energy-momentum tensor is
\begin{eqnarray}
\label{cnN2}
  c^n_N &=& (n-2)\left\{(2n-1)^2 + 2n(n-1)x^2)\right\} \,.
\end{eqnarray}
When $x$ is given by (\ref{xval}) this corresponds to the $q$'th
 unitary minimal model of the $W_{n-1}$-algebra. This relation with
minimal models extends the discussion in \cite{be1}.

Note that in the present case
 of the non-critical string we have the additional freedom of selecting
 the minimal model: the value of $q$ is arbitrary in (\ref{xval}), while
 in (\ref{cnN}) we necessarily obtained $q=N$. This is to be
expected since for $q=N$ we have $c_m=0$ and the theory effectively
reduces to the critical $W$-string. As mentioned in the
 previous footnote, for the non-critical $W$-string
non-unitary minimal models can be considered in the
 same way.

We conclude that in the case of the non-critical string the relation with
 minimal models is not forced upon us, and that
 the non-critical string therefore allows for a much wider
 class of models than the critical string. With a particular choice of
 the parameter $x$ we obtain results similar to those in the
critical case.

It would be very interesting to investigate in further detail the
relations between (critical and/or non-critical) $W$-strings and
minimal models.
The fact that in the non-critical case this relationship can be avoided should
have some significance.
Probably the best way to proceed
is by investigating the cohomology of the different BRST operators in the
``nested'' basis discussed in this paper. An interesting simple example
where the spectrum can be calculated
is provided by taking a non-critical $W_3$-string where the Liouville
sector is realized by just one scalar \cite{be2}.

\vspace{0.5cm}

\leftline{\bf Acknowledgements}

\vspace{.5cm}

The work of E.B.~has been made possible by a fellowship of the
Royal Netherlands Academy of Arts and Sciences (KNAW). The work of
H.J.B.~and S.P.~was performed as part of the research program of the
``Stichting voor Fundamenteel Onderzoek der materie'' (FOM).
Most of the calculations in this paper have been performed by
using the Mathematica package of \cite{th1} for computing operator
product expansions. We would like to thank Jan de Boer, Chris Hull,
Chris Pope, Alex Sevrin and Gerhard Watts and the
participants of the III International Conference on Mathematical
Physics, String Theory and Quantum Gravity (Alushta, 13-24 June, 1993)
for stimulating discussions.

\end{document}